
\overfullrule=0pt

\magnification=1200
\vsize=21.0 truecm
\hsize=14.5 truecm
\hoffset=1.5 truecm
\voffset=1.5 truecm
\baselineskip=12pt

\font\ftit=cmbx10

\parskip=6pt
\parindent=2pc

\font\titulo=cmbx10 scaled\magstep1

\def\section#1{\vskip 0.5truepc plus 0.1truepc minus 0.1truepc
	\goodbreak \leftline{\titulo#1} \nobreak \vskip 0.1truepc
	\indent}
\def\frc#1#2{\leavevmode\kern.1em
	\raise.5ex\hbox{\the\scriptfont0 $ #1 $}\kern-.1em
	/\kern-.15em\lower.25ex\hbox{\the\scriptfont0 $ #2 $}}




\def\Real{{\rm I\!R}}  


\centerline{{\ftit Deterministic Chaos in Tropical Atmospheric Dynamics}
\footnote{*}{This work is supported by CONACyT grant 400349-5-1714E and by the
Association G\'en\'erale pour la Coop\'eration et le D\'eveloppement
 (Belgium).}}

\vskip 1.0pc

\centerline{H. Waelbroeck}

\vskip 0.5pc

\centerline{Instituto de Ciencias Nucleares, UNAM}
\centerline{Circuito Exterior, C. U.}
\centerline{Apdo. Postal 70-543}
\centerline{M\'exico, D.F., 04510 M\'exico.}

\vskip 0.5pc

\baselineskip=22pt

\vskip 0.5pc

\vfill\eject

\vskip 10pc

\centerline { Abstract}

{\noindent
We examine an 11-year data set from the tropical weather station of
Tlaxcala, Mexico. We find that mutual information drops quickly with
the delay, to a positive value which relaxes to zero with a time scale of
20 days. We also examine the mutual dependence of the observables
and conclude that the data set gives the equivalent of 8 variables per day,
known to a precision of $2\%$. We determine the effective dimension
of the attractor to be $D_{eff} \approx 11.7$ at the scale $3.5\% < R/R_{max}
< 8\%$. We find evidence that the effective dimension increases as
$R/R_{max} \to 0$, supporting a conjecture by Lorenz that the climate system
may
consist of a large number of weakly coupled subsystems, some of which have
low-dimensional attractors. We perform a local reconstruction of the
dynamics in phase space; the short-term predictability is modest and
agrees with theoretical estimates. Useful skill in predictions of
10-day rainfall accumulation anomalies reflects the persistence of weather
patterns, which follow the 20-day decay rate of the mutual information.

	\smallskip}

\vfill
\eject

\section{1. Introduction}

 In temperate regions, the large scale approach to meteorology based on
the reduction of fluid equations to a finite grid has been effective in
meeting the demands of agricultural planning and disaster prevention.
This is not the case in tropical regions, where the atmospheric system
is highly sensitive to local ``details'', such as the presence
of a hill or a lake, which are too subtle to be accurately modeled
on a grid system. Often it is not even known which details are important
and which are not. One is tempted to turn to the empirical
approach: to reconstruct the dynamical model not from physical
equations, but from data drawn from observations of the system. This
approach has been applied successfully to scalar time series analysis
(Linsay 1991, Abarbanel et al. 1990, Farmer and Sidorowich 1987),
leading to useful predictions for chaotic systems
for which the underlying equations are not known {\it a priori}. Before
applying this method to meteorology, one must gain a better
understanding of tropical atmospheric dynamics from the point of
view of chaos theory. That is purpose of this article.

 The study of chaos in extended systems, such as atmospheric dynamics,
is attracting a growing interest as the next challenge in dynamical systems
analysis. A key question which must be addressed is whether such systems
can have low-dimensional attractors, so that one could hope to reconstruct
the dynamics in phase space from observations of the system at regular
intervals of time.

 Most attempts at calculating an attractor dimension from weather data
have encountered the difficulty that the data sets are too scarce to trust
that the dimension found is really the correlation
dimension of the attractor. Rather, the scaling graphs generally provide
information only on the large scale structure of the attractor, and
there is no compelling reason to expect that this structure should
be similar at finer scales.
Instead, Lorenz has conjectured that the global climate may be well
represented by a large number of weakly coupled subsystems, some of which
may have low-dimensional attractors [Lorenz 1991]. If the coupling is
sufficiently weak,
the external perturbations would not be detected in
the large-scale reconstruction of the attractor, but would imply a much
larger attractor dimension at smaller scales, reflecting the
greater complexity of the coupled dynamical system. Lorenz based his
conclusion on a study of a model system, which consisted of a
number of identical copies of the set of three modes of the
Oberbeck-Boussinesq equations for the
convection of a fluid heated from below [Lorenz 1963], with
coupling terms connecting the different copies.

 In this article, we will consider Lorenz' conjecture in the
practical setting of a tropical weather data set, and attempt to decide
to what extent a reconstruction of the large-scale structure of the attractor
can be helpful for the purpose of designing prediction models for
extended chaotic systems.

 The data set consists of 19 weather variables measured daily at a single
weather station, located near the town of Tlaxcala, Mexico. The measurements
were performed without interruption or change of equipment over a span of
4015 days (11 years). Holes and flagrant errors in the data set amounted to
less than $0.5 \%$ of the data and were replaced by linear interpolation.
The decision to use 19 variables at a single location, rather than fewer
variables at several locations, was based on evidence that the data had
been compiled most rigorously at the central weather station. Of course
the 19 variables are not all independent; we shall discuss this in more
detail in the next section where we show that the data set carries as
much information as 8 mutually independent variables, or a single
observable over a time span eight times as long.

 The 19 variables are listed in Table 1. The range of each variable
($ = max. - min$.) was chosen to equal $2^{b(i)}$ or a decimal fraction
thereof,
where $b(i)$ is the number of significant bits according to our estimate
of the measurement accuracy. It was then normalized to the unit interval:
$x_k^{(i)} \in [0,1]$, $i = 1, ..., N$, $k = 1,...,n$ ($N = 19$, $n = 4015$)
and the annual cycle was subtracted.  With this
normalization, the greatest possible distance between two points is
equal to one: we will denote the normalized distance by $R / R_{max}$
to recall that it is less than or equal to one (see the precise definition
of the distance function in Sec. 3). The dataset is not available on any
public-domain databank: please contact the author directly for electronic
transfers.

 In Sec. 2, we examine the information content of the data set with
an adaptation of the mutual information method for short data sets.
In Sec. 3, we show that the data reflects a deterministic chaotic
system down to the scale $R/R_{max} = 3.5\%$, and determine the effective
dimension of the attractor in the range $3.5\% < R/R_{max} < 8\%$. A strong
increase in the slope just below this range, but above the scale where the
shortcomings of the data set become noticeable, provides evidence in support
of Lorenz' conjecture on the nature of the climate system. The
theoretical predictability limit of the data set is also estimated, by
two different methods.  In Sec. 4, we describe the results of a local
reconstruction of the dynamics and conclude, based on those results
and the previous data analysis, on how the large scale structure
of the attractor can be used in prediction tasks.

\vfill
\eject

\section{2 Mutual Information Analysis}

 The mutual information between two random variables $x^{(i)}$, $x^{(j)}$
($j \neq i$), with statistics specified by a finite set of events with a
determined
measurement error, is defined as follows. One divides the range of the
variables into a fixed number of intervals (``bins"), such that the size of
each bin is greater than the error and each allowed bin is visited a
statistically significant number of times ($\geq 40$). For two weather
variables with a delay of $\tau$ days, the data set gives us $n - \tau$
events $(x^{(i)}_t,\ x^{(j)}_{t+\tau};\ t = 1, ..., n-\tau)$; a division into
64 bins was found to satisfy both criteria for all variables but $Nos.\ 9, 10$
and $12$ (cloud cover, average and dominant wind directions).

 The probability distribution of a variable is given by the fraction of the
events which fall into each bin: the ``probability of the bin
$\cal{O}_{\alpha}$" is

$$ P(x^{(i)} \in {\cal O}_{\alpha}) = {\# \{ x^{(i)}_t \ (t = 1, ..., n):\
x^{(i)}_t \in
{\cal O}_{\alpha} \} \over n}, \eqno 2.1$$

\noindent where the index $\alpha$ labels the bins and $\#$ is the
cardinality, or number of elements in the set.

 The average information needed to specify the variable $x^{(i)}$, knowing its
probability distribution, is given by

$$I(x^{(i)}) = -<log_2 P(x^{(i)})>, \eqno 2.2$$

\noindent where the average is a sum over $\alpha$ weighted by the probability
$P(x^{(i)} \in {\cal O}_{\alpha})$. One easily checks that the information is
maximal when $P(x^{(i)})$ is homogeneous and $I(x^{(i)})$ is equal to the
number
of bits needed to count the bins: for 64 bins, the maximum information is equal
to 6 bits (Table 1). In general, the more $P(x^{(i)})$ is peaked around a
particular value of $x^{(i)}$, the less information one needs, on average,
to specify the value of $x^{(i)}$: $I(x^{(i)})$ measures ``how surprised one
feels when, knowing the distribution $P(x^{(i)})$, one is told a typical value
of the variable" (Fraser and Swinney, 1983; Fraser, 1988).

 The two-point probability $P(x^{(i)} \in {\cal O}_{\alpha},\ x^{(j)} \in
{\cal O}_{\beta})$ is given by the fraction of events where $x^{(i)}$ falls
into
the bin ${\cal O}_{\alpha}$ and $x^{(j)}$ falls into the bin ${\cal
O}_{\beta}$.
The information needed to specify
two variables is given by

$$I(x^{(i)}, x^{(j)}) = - < log_2 P(x^{(i)}, x^{(j)}) >, \eqno 2.3$$

\noindent where the average is now a sum over $\alpha$ and $\beta$ weighted by
$P(x^{(i)}, x^{(j)})$.

 The information needed to specify two correlated variables jointly is
less than that which is needed to specify each of the variables independently;
the difference measures the degree of mutual dependence of the two variables
and is called ``mutual information". With the notation $I^{(i)} \equiv
I(x^{(i)})$, etc., the mutual information is given by

$$I_m^{(i,j)} = I^{(i)} + I^{(j)} - I^{(i,j)} \geq 0. \eqno 2.4$$

 If one is concerned with the mutual information for two measurements of the
same variable with a delay $\tau$, the relevant events are $\{ x_t^{(i)},
\ x_{t+\tau}^{(i)} \}$. Since $I(x_0^{(i)}) = I(x_{\tau}^{(i)}) =
I^{(i)}$, the mutual information is given by

$$I_m^{(i)}(\tau) = 2I^{(i)} - I^{(i(0), i(\tau))}. \eqno 2.5$$

\noindent Likewise, the mutual information between the variable $x_t^{(i)}$
and the delayed variable $x_{t+\tau}^{(j)}$ is given by

$$I_m^{(i, j)}(\tau) = I^{(i)} + I^{(j)} - I^{(i(0), j(\tau))}. \eqno 2.6$$

\noindent For a chaotic system, $I_m^{(i)}(\tau) \to 0$ and $I_m^{(i,j)}(\tau)
\to 0$ as $\tau \to \infty$, $\forall i,\ j$.

 The mutual information is considered to be the best measure of mutual
dependence of variables, for non-linear systems (Fraser, 1989).

 Unfortunately, the implementation of this method for short data sets
encounters the difficulty that the two-point distribution $P(x^{(i)}, x^{(j)})$
is not accurately defined, since the number of events which fall into each
allowed bin is not statistically significant (insufficient sampling).
The probability distribution comes out rough, which implies that probability
is ``localized" at statistically insignificant peaks in $P(x^{(i)}, x^{(j)})$.
Therefore the two-point information is underestimated and the mutual
information is overestimated. This problem
can be corrected by smoothing the distribution $P(x^{(i)}, x^{(j)})$. We
used the discretized diffusion equation

$$P^{k+1}_{\alpha \beta} = (1-\epsilon)P^k_{\alpha \beta} + \epsilon
\biggl( {P^k_{\alpha \beta +1} + P^k_{\alpha \beta -1} + P^k_{\alpha +1 \beta}
+ P^k_{\alpha - 1 \beta} \over 4} \biggl) \eqno 2.7$$

\noindent which we iterate four times ($k = 1, ..., 4$), so that the final
value of $P_{\alpha \beta}$ is a weighted average of $P$ over
a square of $8 \times 8 = 64$ neighbouring bins, with the strongest weights
assigned to the bins closest to $(\alpha \beta)$. The probability distribution
gradient is set to zero on the boundary, where $\alpha$ or $\beta = 0$ or $65$.
The smoothing parameter $\epsilon$ is chosen
independently for each pair (i, j) so
that $I_m^{(i,j)}(\tau) \to 0$ as $\tau \to \infty$, and for each i, so that
$I_m^{(i)}(\tau) \to 0$. To implement the $\tau \to \infty$ limit, we required
that the average mutual information over the range $\tau = 150, ..., 250$ be
equal to zero; the r.m.s. of $I_m$ over this set of 100 trials gives one
a measure of the degree of precision of the mutual information estimates: an
upper bound $\epsilon < 0.7$ was set on the smoothing parameter, and the cases
where the r.m.s. was greater than 0.05 bits were considered not trustworthy
due to the short data set.

 The mutual information among the 19 variables is given in Table 2. One
notes that in most cases mutual informations are low, between 0 and 0.2 bits,
with a few notable exceptions: for example, the minimum dew point is strongly
dependent on the minimum registered value of the vapor pressure, and likewise
for the corresponding maximum values. This is not surprising, given that
these variables are closely related and reach their extrema more or less
simultaneously. Also the dry bulb temperature is dependent on the minimum
temperature,
cloud cover, minimum vapor pressure and minimum dew point.
The total information at one time step is given by the sum of the informations
of the 19 variables (diagonal terms in Table 2), minus the sum of
the mutual informations $I_m^{(i,j)}$ for
$i < j$, or 70.4 - 23.6 = 46.8 bits, roughly equivalent to 8 variables of
6 bits each. In this counting, one is neglecting the $n-point$ mutual
information, $n \geq 3$.

 The mutual information of each variable as a function of the delay is
given in Table 3.  The mutual information drops to about $10 \%$ after a delay
of one day: this is sufficiently small to justify using successive patterns as
independent axes in a phase space reconstruction, and the total amount of
independent data available for the reconstruction is $8 \times 4015 = 32,120$.

 Some variables have a strong persistence, which is revealed by a
relatively large mutual information at delays $\tau \sim 15$,
particularly the dew point and the vapor pressure, and the dry bulb temperature
and the minimum temperature: this indicates a moderate potential for the
application of statistical prediction models such as the linear autorregressive
models or the low-threshold linear reconstruction model described in Secs. 3
and 4. The slow decline of mutual information with the delay appears to
indicate a strong persistence of some features of weather patterns, which
would be well exploited in such a model. The rainfall shows very
little persistence but has significant mutual dependence with other more
persistent variables. This indicates some potential for medium-range
predictions which would exploit the predictability of these other
variables to give trend predictions for precipitation anomalies (Sec. 4).
This type of predictability through the persistence of another variable could
be quantified as mutual information by computing the three-point
mutual informations.

 The total mutual information as a function of the delay gives one a measure
of how much additional information one gets by adding one day
to a set of $T$ consecutive days of data. A complete description of the state
of the weather system should be such that no new information is gained by
adding more data. Recalling that a single day of data consists of 46.8
bits of information, the information added by introducing a second day is
46.8 - 6.6 = 40.2 bits; the third day adds 46.8 - 6.6 - 4.5 = 35.7 bits, etc.;
the information added per day is represented in Figure 1. The fact that
no additional information is gained after 14 days of data indicates that
the state of the system is well-described if one gives the value of these
variables for 14 consecutive days. This figure is an overestimate: if one
could compute the $n-point$ mutual informations, the total mutual informations
would be larger and the number of days beyond which no new net information
is gained would be less than 14.  We will see below that the embedding
dimension, computed from the scaling graphs, corresponds to 7 consecutive
days of data, and that the local reconstruction of the dynamics gives
optimal predictions for $T_e = 10 - 14$. The mutual consistency of these
results is encouraging.

\vfill
\eject

\section{3. Phase Space Reconstruction}

 The phase space reconstruction is based on the following idea. One assumes
that the system can be modeled by $d$ first-order differential equations;
we will see below how one verifies the validity of this assumption and
determines $d$. Any set of $d$ independent
measurements of the state of the system provides a complete
set of initial data which determines the trajectory uniquely.
In the phase space reconstruction method, one begins by choosing $d$
observables to provide a convenient representation of the state of
the system in $\Real^d$. These
can be measurements of a single physical variable at regular intervals
of time, as in the method of delays, or measurements of $d$ different
variables when another variable equals a predetermined
value (Poincar\'e section). Here we will take the 19 available variables
for $T$ consecutive days of data ($d = 19 \times T$). Some of them are
minimum or maximum values, others are averages over a 24-hour period;
all are well-defined functions of the continuous-time
physical variables over non-overlapping intervals of time. Following the
assumption that the dynamics is determined by first-order differential
equations, these induce recursion relations
for the evolution of the observables from one 24-hour period to
the next. The phase space reconstruction proposes (1) to examine the
geometry of the reconstructed attractor (fractal dimension, etc.) and
(2) to find an approximation of these recursion relations.

 We define a point in phase space as a set of $19 \times T$ variables
corresponding to $T$ consecutive days of observations. The data set
then gives one $4015 - T$ points with which we shall attempt to reconstruct
the attractor in the embedding phase space: $\buildrel \to \over x_k =
\{ x^{(i)}_{k-t}: \ i = 1, ..., N;\ t = 0, ..., T-1 \}$. We will use the
arrowed vector notation for phase space points, and bold-face letters
for a single day of observations: ${\bf x}_k \in \Real^N$. The distance
between two points in phase space is taken to be the
normalised Manhattan distance

$$d(\buildrel \to \over x_k, \buildrel \to \over x_l) = {1 \over N}
\sum_{i=1}^N \sum_{t=0}^{T-1} \alpha(t) \mid x^{(i)}_{k-t} - x^{(i)}_{l-t}
\mid \eqno 3.1$$

\noindent where $\alpha(t)$ is a partition of the identity for $t = 0, 1, ...,
T-1$, which in this
section will be taken to be uniform: $\alpha(t) = 1/T,\ \forall t$. Since
the range of each variable $x^{(i)}$ was normalised, one easily checks that
$d(\buildrel \to \over x_k, \buildrel \to \over x_l) \leq 1$, $ \
\forall k,l$.

 To determine the correlation dimension, one computes the number of pairs
of points with distance $d(\buildrel \to \over x_k, \buildrel \to \over x_l)
< \rho$: $N(\rho)$ (Grassberger and Procaccia, 1983). If $T=T_e$ is chosen
so that $\Real^{N\times T_e}$ embeds the attractor of dimension $D_a$, then
$N(\rho) \sim \rho^{D_a}$. If $T$ is too small for a proper embedding, $T<T_e$,
then $N(\rho) \sim \rho^{D(T)}$, where $D(T)$ is the dimension of the
attractor projected onto the first $T\times N$ coordinates. The dimensions
$D(T)$ are determined from the slopes of logarithmic plots of $N(\rho)$,
by linear regression (Figure 2). If there is a deterministic
dynamical rule, the graph of $D(T)$ reaches an upper
bound $D = D_a$ when $T = T_e$. There are several obstacles
which can inhibit a linear ``scaling region" in the graph of $log(N)$ vs.
$log(\rho)$.

\noindent 1. Noise. A stochastic signal superimposed on the deterministic
signal can disperse the points along directions orthogonal to the $D_a$-
dimensional attractor; this is evidenced by a steeper slope of the
logarithmic graph. In the case of our data set, measurement error
can be estimated to be less than
$2 \%$ of the range of each variable; if we assume that the signed
error is equally distributed on both sides of the origin, the measurement
error in the Manhattan distance between points in $\Real^{N\times T}$ is
of the order of (Feller 1968)

$$\epsilon_m \sim {2 \over \sqrt{T\times N} }\ \% . \eqno 3.2$$

\noindent For systematic measurement errors the random-sign argument is
not valid and one must consider the upper bound on the error,
$\epsilon_m < 2 \%$.

\noindent 2. Lack of data. For a short data set, at small $\rho$ the total
number of pairs of points with distance less than $\rho$ becomes too small
to be considered statistically significant and one should require that

$$N(\rho) \geq 50 . \eqno 3.3$$

\noindent In our case, the graphs of $N(\rho)$ reveal that this places a
lower bound $\rho \approx 3 - 4 \%$, depending on $T$, below which
the points of the graphs scatter randomly due to the bad statistics.  Since
this lower bound is relatively large, one cannot safely assume that, for an
infinitely long data set, the scaling region would continue with the same
slope all the way down to $\rho \to 0$. This problem is frequently
encountered in climate and weather data sets (see Grassberger 1986,
Lorenz 1991, and references therein) and implies that the
correlation dimension, defined as the effective dimension
in the limit $\rho \to 0$, cannot be computed. One should then consider the
dimension estimates as effective dimensions at a certain scale. As Lorenz
has suggested, it is possible that this effective dimension be a good
approximation of the correlation dimension of a simpler local sub-system
which is more or less decoupled from the global climate system: at lower
values of $\rho$, the coupling with external variables would become relevant,
leading to a much higher correlation dimension. We will describe evidence in
support of Lorenz' conjecture below.

\noindent 3. Edge effect. For large values of $\rho$, comparable to the
size of the attractor, the number of pairs saturates when a sphere of radius
$\rho$ centered at a typical point on the attractor extends beyond the
edge of the attractor, in an empty region of phase space: the maximum possible
number of pairs taken from $n$ data points is $n^2$. The scaling region
therefore ends below this saturation point, and

$$N(\rho) << n^2. \eqno 3.4$$

\noindent To see exactly where saturation becomes significant,
one can only draw the complete graph
of $N(\rho)$ and observe at which point saturation occurs. From Figures 2-4,
one sees that saturation becomes significant for $N(\rho) \geq n^2/100$.

\noindent 4. Scale-dependent dimension. Some attractors have different
apparent dimensions at different scales; in some cases even the small-scale
limit of the effective dimension is not well-defined and the correlation
dimension does not exist, as for example with Zaslavskii's map, where $D_a$
alternates between $D_a \approx 1$ and $D_a \approx 1.6$
in the limit $\rho \to 0$ (Grassberger and Procaccia, 1983). If the effective
dimension varies with $\rho$ in the region between the lower
bounds (subsections 1 and 2) and the upper bound (s.s. 3), it may be impossible
to find a physically significant scaling region.

 In our case, a clear scaling region is found up to $T = 10$ (Figure 3)
and to a lesser extent at $T = 14$ (Figure 4). The scaling regions are
taken to begin above $N(\rho)/n = 0.01$ $(N(\rho) \approx 40)$ and extend
to larger $\rho$ with the limitation that the linear regression
coefficient be greater than or equal to 0.999. The slopes of the scaling
regions (Figure 5) give $D(2) = 8.5$, $D(4) = 9.7$, $D(7) = 11.12$, $D(10)
 = 11.67$ and $D(14) = 11.68$,
indicating an effective attractor dimension $D_{eff} = 11.7$ at the scale
$0.04 < \rho < 0.08$.

 At larger scales ($0.08 <\rho < 0.12$), one notes a slight $increase$
in the effective dimension, to $D_{eff} \sim 14$, although the scaling
region is not sufficient to give an accurate estimate of the slope. This
contrasts with the decrease in the slope with saturation, at $\rho > 0.12$.

 At the other end of the correlation graph, the first few points for which we
can consider that the statistics is satisfactory suggest a steeper slope,
particularly for larger values of $T$ (Figures 4, 6); this hints that the
effective dimension of the dynamics at small scales may be greater than in the
scaling region, perhaps due to a weak coupling to external variables as
suggested by Lorenz' work (Lorenz, 1991). Let us assume that this external
coupling can be approximated by a random variable with variance $\sigma^2$
and mean zero, added on to each phase space variable. The
average squared distance between two points in phase space
would then be $\rho^2 = \rho'^2 + 19T\sigma^2$,
where $\rho'$ is the average without noise. For example,
one can determine from Figures 2-4 the average distance $\rho_{200}$ such
that there are $N(\rho_{200}) = 200$ pairs of data points with distance less
than $\rho_{200}$. From $N(\rho_{200}) \approx 0.04$ at $T = 7$ and
$\rho_{200} \approx 0.05$ at $T = 14$ one finds $19\sigma^2 \approx
1.3 \times 10^{-4}$. This allows one to compute $\rho_{200} \approx 0.065$
at $T = 29$, in agreement with Figure 6. Although this is by no means a
formal argument, the accuracy of the prediction for $T = 29$ indicates
that one may be well justified to interpret the correlation graphs as
follows: the correlation graphs indicate the existence of a deterministic
system
with attractor dimension $D_a = 11.7$ coupled to external perturbations,
which can be reasonably well modeled as a random term added on to each
variable, with variance $\sigma^2 \approx 1.3 \times 10^{-4}$ and mean zero.

 If the deterministic signal is a representation of the local
atmospheric dynamics and the random part is a representation of external
perturbations, then one might think that a fully deterministic model
could be obtained if one included in the phase space the so-called
``external" variables, perhaps by including measurements of other
variables besides the $N = 19$ variables of the data set. A lower bound on the
dimension of the attractor of the coupled system can be derived
from the slope of the correlation graph at low values of $\rho$;
this would give an effective dimension in excess of $D_{eff} \approx 35$
for $\rho \leq 0.04$ (T = 7) or $\rho \leq 0.05$ (T = 14). Such a
large attractor dimension indicates that a reconstruction of the
phase space would require a very large number of data points, corresponding
to daily observations over at least $10^{35/11.7}$ years just to
reconstruct the attractor at the same very low resolution with which we
reconstructed the 11.7-dimensional attractor. The approach suggested
in the previous paragraph, to treat the perturbations as random noise,
stands out as more realistic.

 Whether or not the speculations of the previous two paragraphs stand up
to future scrutiny, one can surely not assume that the effective
dimension $D_{eff} \approx 11.7$ is valid at all scales, or that it is
a valid estimate of the correlation dimension of the attractor, defined
as the effective dimension in the limit $R \to 0$.

 At the scale $0.04 < \rho < 0.08$, the effective dimension $D_{eff} \approx \
11.7$ appears to be trustworthy: each point in the scaling region is based on a
large number of pairs, so that $N(\rho)/n^2$ is insensitive to the
size of the data set (this has been checked explicitly by further
reducing the data set), and is well below the saturation,
as is evidenced by the fact that the slope does not drop off until much
larger values of $N(\rho)$. Some numerical experiments lend further confidence
to the effective dimension result:

\noindent 1. The sensitivity to the choice of delay was
checked by averaging the data by sets of three days and thereby reducing the
length of the data set to $n/3$ days; the same attractor dimension was
recovered.

\noindent 2. The sensitivity to the choice of variables was examined
by considering only the first 6 variables for each day, instead of 19.
Again, the same attractor dimension was found. A further reduction, to
the first 2 variables, led to an underestimation of the effective
dimension, $D_{eff} \approx 11.1$, which indicates that these two
variables were not perfectly coupled to the system as a whole. Although
this result may surprise strong believers in the so-called ``Takens rule",
following which any one variable should reflect the dynamics of the
entire system through its interactions with all others, in practice
if the interaction is too weak then they are detected only
in the fine structure of the attractor, below the scale $\rho = 0.04$.
This implies that if one is going to use a single variable in the
correlation method, then it is important to choose one which is strongly
coupled to the others, as noted by Lorenz (Lorenz 1991).

\noindent 3. Finally, we performed the standard check of generating
an artificial data set with the same statistics but no deterministic
dynamics, and found that the upper bound $D(T) \leq D_{eff}$ disappears. We
generated such a data set by randomly mixing the $n = 4015$ patterns,
and found that the graph $D(T)$ increases monotonously, passing $D = 34$
at $T = 14$.

 The largest Liapunov exponents can be estimated by considering the
average rate of divergence of nearby trajectories (Briggs 1990, \
Gade and Amritkar 1990, Ellner et al. 1991, Galbraith 1992, Wales 1991,
Zeng, Eykholt and Pielke 1991). For each available
data point in the phase space, $\buildrel \to \over x_k \in \Real^{T_e
\times N}$, the nearest neighbour is identified and its evolution is
compared to that of the original pattern, step by step. Here we
excluded near-neighbors that are also neighbors in time (measurements
within 20 days of each other) since their trajectories do not diverge.
The error at each time step is averaged over the patterns. The error grows
rapidly with a time scale roughly equal to 2 days, then reaches a slowly
rising upper bound which is significantly lower than the average distance
between randomly chosen points (Figure 7). The upper bound indicates
a potential for longer-term predictability based on the persistence of
weather patterns. The doubling of the error in two days would indicate
a Liapunov exponent $\Lambda_1 \sim {1 \over 2} ln(2)$, although the data set
is clearly insufficient to consider this a serious estimate.

 A quantitative estimate of the theoretical predictability limit can
be obtained as a byproduct of the correlation method (Fraedrich 1987,
Fraedrich and Leslie 1989, Nese 1989).

$$T_{pre} = {ln(N(\rho; T)) - ln(N(\rho; T+m)) \over m}. \eqno 3.5$$

\noindent From the slope of the graph of $N(\rho; T)$
as a function of $T$, with $\rho = 0.05$, one finds $T_{pre} = 1.83$ days
(Figure 8). Another estimate is given by the slope of the graph $D(T)$
in the linear region previous to the saturation at $D(T_e) = D_{eff}$. A
linear regression (Figure 9) gives $T_{pre} = 2.1$ by this criterion.
The similarity between these estimates, and the consistence with the
divergence rates discussed in the previous paragraph, hints that they
may be valid in spite of the very short data set. The prediction
error at time $T$ would then be $\epsilon(T) \approx \epsilon_0
\ e^{T/T_{pre}}$ for small $T$, where $\epsilon_0$ is the scale at
which the reconstruction of the attractor is meaningful, $\epsilon_0
\sim 4 \%$. Note that the predictablility limit falls well short of the
2-3 week limit of large scale weather patterns studied in General
Circulation Models (GCM's). This confirms the claim made in the introduction,
that the high sensitivity of local atmospheric dynamics in the tropics
invalidates the grid approach of the GCM's for aspects of the weather
that are sensitive to the local variables, such as tropical precipitation.
Grid-reductions of the fluid equations make sense only if the system
is sufficiently stable that the inevitable simplifications in the
equations and the grid reduction itself do not lead to a numerically
unstable code. For highly unstable systems,
the attractor reconstruction method is an obvious candidate to guarantee
numerical stability without over-simplifying the dynamics. Unfortunately
the high dimensionality of the attractor severely limits the accuracy
of the phase space reconstruction method in this particular example, as
we shall see shortly.

\vfill
\eject

\section{4. Local Reconstruction of the Dynamical Map}

 In the previous section, we saw that the state of the weather system
can be described by a set of $T_e \approx 10$ consecutive days of
measurements, corresponding to one point in the phase space
$\Real^{T_e \times N}$. Here, we will be concerned with the reconstruction
of the dynamics (Farmer and Sidorowich 1987, 1988, Abarbanel et al. 1990,
Giona, Lentini and Cimagalli 1991, Gouesbet 1991, Linsay 1991, Tsonis 1992,
Elsner and Tsonis 1992). Given an initial point $\buildrel \to \over x_0$
in this phase space, we wish
to compute its orbit. The dynamical evolution integrated over one time
step is a map ${\bf f}: \Real^{T_e \times N} \to \Real^{T_e \times N}$; we
will denote by ${\bf f}_L$ the local reconstruction of this map in the
vicinity of $\buildrel \to \over x_0$. A generalization of the method of
analogues will be used (Lorenz 1963, 1969, van den Dool 1989, Toth 1989): the
evolution of the initial point $\buildrel \to \over x_0$ will be based
on the data points $\buildrel \to \over x_k$ closest to $\buildrel \to
\over x_0$ and their known evolution one day ahead, ${\bf x}_{k+1} =
{\bf f}(\buildrel \to \over x_k)$

 We will use the distance (3.1), with weights $\alpha(t)$ decreasing
linearly with the delay so that $\alpha(T_e-1) = \alpha(0) / 2$.

$$\alpha(t) = {4(T_e-1)-2t \over 3T_e(T_e-1)} \eqno 4.1$$

 We will consider the data points $\buildrel \to \over x_k$ such that
$d(\buildrel \to \over x_k, \buildrel \to \over x_0) < \eta$, where
$\eta$ is the radius of the ball around $\buildrel \to \over x_0$ where
we wish to determine the local map ${\bf f}_L$. The reconstructed
map is given by

$${\bf f}_L(\buildrel \to \over x_0) = {{\sum_{k=1}^n
\lambda \biggl( \eta - d(\buildrel \to \over x_k, \buildrel \to \over x_0)
\biggr)  {\bf x}_{k+1}} \over {\sum_{k=1}^n
\lambda \biggl( \eta - d(\buildrel \to \over x_k, \buildrel \to \over x_0)
\biggr)}} \eqno 4.2$$

\noindent where $\lambda$ is a threshold-linear weight function,

$$\lambda(x) = x \ \theta(x) \eqno 4.3$$

\noindent for a local linear reconstruction ($\theta$ is the step function).

 Thus, the dynamical model (4.2) is a weighted average of
${\bf x}_{k+1} = {\bf f}(\buildrel \to \over x_k)$ over the
points $\buildrel \to \over x_k$ in an $\eta$-ball centered at
$\buildrel \to \over x_0$, weighted by $\eta - d(\buildrel \to \over x_k,
\buildrel \to \over x_0)$.

 For low values of $\eta$, this model functions as an associative
memory to recall a temporal sequence of patterns, with a storage capacity that
can be computed analytically using statistical methods developed for neural
networks (Zertuche et al., 1994a,b). Here, we are interested in supercritical
values of $\eta$, where the learned sequence is unstable and the model acquires
positive Liapunov exponents, like the physical system being modeled.

 The prediction model has two parameters which must be determined by
optimization: the radius $\eta$, and the number of time
steps which determine a point in phase space: $T_e$.

 Based on the previous results on mutual information estimates and the
application of Grassberger and Procaccia's algorithm, one would expect
that $T_e$ should be sufficient to give a proper embedding of the
attractor, $T_e \geq 7$, and $\eta$ should be such
that a typical initial point has a sufficient number of neighbours with
distance less than $\eta$ for a meaningful linear interpolation. For the
given embedding dimension, the number of points should be $N(\eta) \approx
500$;
which would require that $\eta \approx 0.1$.
Unfortunately, this is not far below the saturation point ($N \leq 4000$),
so one cannot expect a very fine reconstruction of the dynamics: the true
dynamical map over such a large $\eta$-ball is not likely to be
approximately linear, and therefore ${\bf f}_L(\buildrel \to \over x_0)
\neq {\bf f}(\buildrel \to \over x_0)$.

 The optimization was performed by removing the last year of data and
using it to extract a statistical sampling of initial test-points $\buildrel
\to \over x_0$. The skill for one-day lead forecasts was found to
be optimal for $T=14$, $\eta=0.1$, where

$$skill = 1 - << d_1( {\bf f}_L(\buildrel \to \over x_0), {\bf x}_{k+1}) >>.
\eqno 4.4$$

$$d_1({\bf x}_k, {\bf x}_l) = {1 \over N} \sum_{i=1}^N \mid
x_k^{(i)} - x_l^{(i)} \mid \ . \eqno 4.5$$

 The skill of the optimal model ($T = 14$, $\eta = 0.1$) is

$$ Max(skill) = 0.919 \eqno 4.6$$

\noindent which compares to 0.904 for persistence, 0.905 for
the seasonal average and 0.865 for random pattern selection. More details
on the results of the prediction code are being published separately
(Waelbroeck et al. 1994).

 At optimum, we also measured the skill in the prediction of the anomaly
of rainfall accumulations over a 10-day period, a variable chosen for
its relevance in agriculture: the skill was equal to $64 \%$ of the
variance of the observed anomaly over the 11-year span of measurements, where
the
anomaly of a variable at a given date is defined as the difference between the
value of the variable and its seasonal average, evaluated over all
available data within 10 days of this date, for each of the
11 years of data (an average over 231 samples).

 The 10-day rainfall predictions for all of 1992 (the last year in the
data set) are compared to the seasonal average and to the observed
precipitations in Figure 10. Although the predictions clearly do not
follow the observations very closely, the unusual early beginning
of the rain season was predicted by the model.

\vfill\eject

\section{5. Summary}

 In summary, we have shown that the effective dynamics of a tropical
weather system, at a scale corresponding to $3.5 - 8 \%$ of the range
of the variables, reflects an underlying deterministic dynamics
with an attractor of dimension $D_a \approx 11.7$. The local linear
structure of the attractor indicates a theoretical predictability of
about 2 days for linear reconstructions of the dynamics.

 A sharp increase in the effective dimension at smaller
scales, $R/R_{max} < 4 \%$, may indicate that external perturbations of the
local
system become significant, in accordance with Lorenz' conjecture that the
climate system consists of a number of weakly coupled subsystems whose
attractors can be detected with the Grassberger-Procaccia algorithm.

 A simple reconstruction of the dynamics was proposed, which showed some skill
in 10-day predictions: this long-term predictability appears to
reflect the persistence of weather patterns, which would also cause the slow
decline of the mutual information as a function of the delay.

 Altogether, three data analysis methods were applied and gave reasonably
coherent results for the following characteristics of the system.

\noindent 1. Embedding dimension: The mutual information analysis gave
$T_e = 14$, the scaling graphs led to $T_e = 7$ and the optimization of
the local reconstruction model gave $T_e = 14$.

\noindent 2. Short-term predictability, limited by $\Lambda_1$: The mutual
information analysis indicates that $T_{pre} \leq 1$ day, the scaling
graphs gave $T_{pre} \leq 2$ days and the local reconstruction model,
$T_{pre} \leq 1$ day.

\noindent 3. Long term predictability, from the persistence of weather
patterns: After a quick initial drop, the mutual information fades out
more slowly with a time scale of 20
days. Also, the distance of nearby trajectories quickly reaches a maximum
distance $\rho \approx 0.1$ after which it increases more slowly, with a time
scale of about 20 days. Finally, the prediction model shows that
useful skill can be extracted for 10-day predictions.

 The general coherence of these results lends some confidence that they
reflect genuine physical properties of the dynamical system. There appears
to be three effective dynamics, at three different scales.

\noindent 1. At very fine scales, one has a very complex system involving
the local system and its couplings to the external variables, with an
effective dimension of the attractor $D_a \geq 35$.

\noindent 2. At the scale $0.04 < \rho < 0.08$, the local weather system
dominates, with an effective attractor dimension $D_a \approx 11.7$, a
Liapunov exponent $\Lambda_1 \geq {1 \over 2} ln(2)$ and a predictability
limited to a couple of days.

\noindent 3. At the scale $\rho \approx 0.1$, one has general
features of weather patterns which follow a much softer dynamical
evolution, characterised by long predictablity times, of the order of
20 days.

 The situation is similar to that encountered in linear regression
forecasts, where a short-term predictability due to a first-order
Markov process is followed by a longer ``tail of autocorrelation"
which allows prediction of monthly means well beyond what can be
expected from the first-order process (Gutzler and Mo 1983, Horel
1985, Trenberth 1985, Van den Dool, Klein and Walsh 1986, Roads and
Barnett 1984, Strauss and Halem 1981)\footnote{*}{We
thank the referee for this comment}.
Of course in this article the short term predictability is modeled
by a non-linear rule based on 14 consecutive days of data, but a
similar ``tail of mutual information" is likely to be responsible for the
quality of the predictions at medium-long range, as anticipated in
Section 2.

 The external perturbations only affect the system progressively,
reaching the amplitude $\rho = 0.05$ after 14 days; this indicates
that the short-term predictability could be improved and extended
several days, if one could improve on the resolution with which the local
dynamics is reconstructed: this would
require improving the use of the available local data to better define
the structure of the attractor near the scale $R/R_{max} \approx 4 \%$.

 Beyond that point, any further improvement in prediction skill would
require introducing the large scale external perturbations
from general circulation
models. This last task requires a methodology for determining what
aspects of large-scale weather patterns are most relevant to the local
system. We hope that the preliminary work which we have described in
this article has laid the first stones towards a new approach to small-scale
meteorology, which is custom-designed to predict local, chaotic
phenomena.

\vfill
\eject

\section{Acknowledgments}

We would like to express our gratitude to Tom\'as Morales of the Center for
Atmospheric Sciences and to members of the Tlaxcala regional center
of the {\sl Comisi\'on Nacional del Agua} for access to the data, and thank
S. Orozco, J. Jim\' enez, R. Orea, M. V\' azquez and A. Meneses for
the preparation of the data set. We also gladly express our gratitude
to the {\sl Comit\'e de Superc\'omputo} of the UNAM for access to the
Cray YM-P4. Finally, we are indebted to C. Duqu\'e,
director of the {\sl Association pour le D\'eveloppement par la
Recherche et l'Action Integr\'ees} and the Belgian Ambassador to Mexico,
for their support of our project.

\vfill\eject

\noindent REFERENCES

\noindent Abarbanel, H. D. I., Brown, R. and Kadtke, J. B., 1990: Prediction in
chaotic nonlinear systems:  Methods for time series with broadband Fourier
spectra. Phys. Rev. 41A 	N!4, 1782.

\noindent Anthos, R. A., 1991: Predictability of mesoscale meteorological
phenomena. In 	Predictability of Fluid Motions.  G. Holloway and

\noindent B. J. West, Eds. American 	Institute of Physics, pp. 247-270.

\noindent Briggs, K., 1990:  An improved method for estimating Liapunov
exponents of chaotic time 	series. Phys. Lett. A 151, 27.

\noindent Ellner, S., Gallant, A. R., Mc.Caffrey, D., and Nychka, D., 1991:
Convergence rates and 	data requirements for Jacobian-based estimates
of Lyapunov exponents from data. 	Phys. Lett. A 153, 357.

\noindent Elsner, J. B., and Tsonis, A. A., 1991: Do bidecadal oscillations
exist in the global 	temperature record.  Nature 353, 551.

\noindent Elsner, J. B., and Tsonis, A. A., 1992: Nonlinear prediction, chaos,
and noise.  Bull. Am. 	Meteo. Soc. 73 N!1, 49.

\noindent Essex, C., Lookman, T., and Nerenberg, N. A. H., 1991:  The climate
attractor over very 	short time scales.  Nature 326, 64.

\noindent Farmer, J. D., and Sidorowich, J. J., 1987: Predicting chaotic time
series.  Phys. Rev. 	Lett. 59, 845.

\noindent Farmer, J. D., and Sidorowich, J. J., 1988: Exploiting chaos to
predict the future and 	reduce noise.  Los Alamos preprint  LA-UR-88-901.

\noindent Feller, W., 1968: An introduction to probability theory and its
applications. Wiley,
	New York., pp. 244.

\noindent Fraedrich, K., 1986: Estimating the dimensions of weather and
climate attractors.  J. 	Atmos. Sci. 43  N! 5, 419.

\noindent Fraedrich, K., 1987: Estimating weather and climate predictability
on attractors.  J. Atmos. 	Sci. 44 N! 4, 722.

\noindent Fraedrich, K., and Leslie, L. M., 1989: Estimates of cyclone track
predictablility I: 	Tropical  cyclones in the Australian region.
Q. J. R. Meteorol. Soc. 115, 79.

\noindent Fraser, A. M. and Swinney, H. L., 1983: Independent coordinates for
strange attractors
	from mutual information. Phys. Rev.  A 33, 1134.

\noindent Fraser, A. M., 1988: Phase space reconstructions from time series.
Physica  D34, 391.

\noindent Fraser, A. M., 1989: Information and entropy in strange attractors.
I.E.E.E. Transactions
	on  Information Theory  35, 245.

\noindent Frison, T., 1990: Predicting nonlinear and chaotic systems behavior
using neural 	networks.  J. Neural Net. Comp. 2, 45.

\noindent Gade, P. M., and Amritkar, R. E., 1990:  Characterizing loss of
memory in a dynamical 	system. 	Phys. Rev. Lett. 65 N! 4, 389.

\noindent Galbraith, J. W., 1992:  Inference about trends in global temperature
data. Clim. Change 	22, 209.

\noindent Ghil, M., and Vautard, R., 1991:  Interdecadal oscillations and the
warming trend in global 	temperature time series.  Nature 350, 324.

\noindent Giona, M., Lentini, F., and Cimagalli, V., 1991: Functional
reconstruction and local 	prediction of chaotic time series.
Phys. Rev. A 44, 3496.

\noindent Gouesbet, G., 1991:  Reconstruction of the vector fields of
continuous dynamical systems 	from numerical scalar time series.
Phys. Rev. A 43, 5321.

\noindent Grassberger, P., and Procaccia, I., 1983:  Characterization of
strange  attractors.  Phys. 	Rev. Lett. 50, 346.

\noindent Grassberger, P., 1986: Do climatic attractors exist? Nature  323,
609-612.

\noindent Gutzler, D. S. and Mo, K. C., 1983: Autocorrelation of northern
hemisphere geopotential
	heights. M. Weath. Rev.  111(1), 155-164.

\noindent Hammel, S. M., 1990:  A noise reduction method for chaotic systems.
Phys. Lett. A 148 	N! 8, 421.

\noindent Henderson, H. W., and Wells, R., 1988: Obtaining attractor dimensions
from 	meteorological time series.   Adv. in Geophys. 30, 205.

\noindent Hense, A., 1987: On the possible existence of a strange attractor for
the southern 	oscillation.   Beitr. Phys. Atmosph. 60, 34.

\noindent Horel, J. D., 1985: Persistence of the 500-MB height field during
northern hemisphere
	winter.  M. Weath. Rev.  113(11), 2030 - 2042.

\noindent Kaplan, D. T., and Glass, L., 1992:  Direct test for determinism in
a time series.  Phys. 	Rev. Letters  68 N! 4, 427.

\noindent Katz, R. W., 1992:  Statistical evaluation of climate experiments
with general circulation 	models:  A parametric time series modelling
approach.  J. Atmos. Sci. 39, 1445.

\noindent Linsay, P. S., 1991: An efficient method of forecasting chaotic
time-series using linear 	interpolation.   Phys. Lett. A 153 N!6, 353.

\noindent Lorenz, E. N., 1963:  Deterministic nonperiodic flow.  J. Atmos.
Sci. 20, 130.

\noindent Lorenz, E. N., 1969: Atmospheric predictability as revealed by
naturally occurring 	analogues.   J. Atmosph. Sci. 26, 636.

\noindent Lorenz, E. N., 1991: Dimension of weather and climate attractors.
Nature  353, 241.

\noindent Nese, J. M., 1989:  Quantifying local predictabiliy in phase space.
Physica D 35, 237.

\noindent Nicolis, C., and Nicolis, G., 1984:  Is there a climate attractor?
Nature 311, 529.

\noindent Pandit , S. M., and Yu, S. M., 1983:  Time series and system analysis
with applications 	(Wiley, New York),  272 pp.

\noindent Roads, J. O. and Barnett, T. P., 1984: Forecasts of the 500-MB height
using a
	dynamically oriented statistical model. M. Weath. Rev. 112(7), 1354 - 1369.

\noindent Sharifi, M. B., Georgekakos, K. P., and Rodriguez-Iturbe, I., 1990:
Evidence of 	deterministic chaos in the pulse of storm rainfall.
J. Atmos. Sci. 47, 888.

\noindent Strauss, D. M. and Halem, M., 1981: A stochastic-dynamical approach
to the study of
	natural variability of the climate. M. Weath. Rev.  109(3), 407 - 421.

\noindent Toth, Z., 1989:  Long-range weather forecasting using an analog
approach.  J. Climate 2, 	594.

\noindent Tsonis, A. A., 1992:  Chaos.  From theory to applications.
(Plenum Press, New York),
	274 pp.

\noindent Tsonis, A. A., and Elsner, J. B., 1988:  The weather attractor over
very short time scales. 	Nature 333, 545.

\noindent Tsonis, A. A., and Elsner, J. B., 1989:  Chaos, strange attractors
and weather.  Bull. 	Amer. Meteo. Soc. 70, 14.

\noindent van den Dool, H. M., Klein, W. H. and Walsh, J. E., 1986:  The
geographical distribution
	and seasonality of persistence in monthly mean air temperatures over
the United
	States. Mon. Weath. Rev.  114(3), 546 - 560.

\noindent van den Dool, H. M., 1989:  A new look at weather forecasting through
analogues.  Mon. 	Wea. Rev. 117, 2230.

\noindent Vautard, R., and Ghil, M., 1989:  Singular spectrum analysis in
nonlinear
dynamics, with 	applications to paleoclimatic time series.  Physica  D 35, 395.

\noindent Waelbroeck, H. et al., 1994: Prediction of tropical rainfall by local
phase space
	reconstruction. Mexico Preprint ICN-UNAM-94-02.

\noindent Wales, D. J., 1991:  Calculating the  rate of loss of information
from
chaotic time series by 	forecasting.  Nature  350, 485.

\noindent Zeng, X., Eykholt, R., and Pielke, R. A., 1991: Estimating the
Lyapunov
exponent 	spectrum from short time series of low precision.
Phys. Rev. Letters  66, 3229.

\noindent Zertuche, F. et al., 1994a: Storage capacity of a neural network with
state-dependent
	synapses. J. Physics  A27, 1575 - 1583 .

\noindent Zertuche, F. et al., 1994b: Recognition of temporal sequences of
patterns using state-
	dependent  synapses. J. Physics  A, in press.

\vfill\eject

\noindent FIGURE CAPTIONS

\noindent Table 1  The 19 variables observed each day are given, as well as
the number of relevant bits indicating the measurement accuracy, and
the information calculated by partitionning the range of each variable
into 64 subintervals.  The maximum and minimum value of each variable
prior to normalization is also given. Atmospheric pressure is corrected
for elevation.

\noindent Table 2  The mutual information of each pair of observables at the
same
time step is given; again the interval [0,1] is divided into 64 subintervals,
so that the maximum possible information is equal to 6 bits. Note that the
diagonal terms represent the information of each variable individually.
The sum of all the terms on the diagonal gives the information needed to
specify the 19 variables independently, while the sum of all terms below
the diagonal gives the total amount of mutual information among them. The
difference between these two quantities is equal to the information needed
to specify the 19 variables together.

\noindent Table 3  The mutual information of each variable is given as a
function
of the delay, as well as the total over all variables. Some variables have
a relatively large persistence even for delays of 14 days, while for other
variables the mutual information drops very quickly with the delay. The
total mutual information at a delay t = 1 day is less than 10
total information in one day of data, so one can consider that consecutive
days provide reasonably independent coordinates for a phase space
reconstruction.

\noindent Figure 1  The additional information needed to specify the variables
at
day T+1, knowing these variables at days number 1, 2, ..., T, is given as
a function of T.  No new information is gained by extending the phase space
vector beyond 14 consecutive days of data. This indicates that the dynamics
is deterministic with an embedding dimension corresponding to Te E 14.

\noindent Figure 2  The dimension D(T) of the attractor's projection on a
19T-dimensional subspace of the embedding space is given, as a
function of T.  A plateau is clearly reached at T=7, indicating an
effective dimension of the attractor Da E 11.7.

\noindent Figure 3  The scaling graph gives the number of pairs of data points
in phase space with distance less than r, as a function of r, if we
use a slightly redundant embedding with 10 consecutive days of data.
One distinguishes four regions at increasing values of r: at the lowest
values there appears to be a tendency for the slope to increase, but
this occurs at values of N too low to trust the statistics. Next comes
the scaling region, with a slope equal to Da E 11.67. One then enters a
non-linear region where the slope increases at first, to about 14, before
it begins to decrease as a consequence of the saturation, when r becomes
comparable to the size of the attractor.

\noindent Figure 4  The scaling graph for T = 14 consecutive days of data
displays
the same features as for T = 10, although the scaling region with a slope
equal to Da E 11.68 is narrower. The increase in the slope at low values
of r is more significant. This may reflect a higher effecive dimension of
the system coupled to external variables, as suggested by Lorenz' work;
a time of 14 days would then appear to be sufficient to affect the local
system up to the statistically significant scale rE 0.055.

\noindent Figure 5  The number of pairs within a ball of radius r, N(r), scales
like a power of r.  The bilogarithmic plot gives a straight line which
can be determined by linear regression. The scaling graphs are
represented for four different values of the embedding dimension,
correslonding to T = 2, 4, 7 and 10.  The projected dimension D(T) of
the attractor on the chosen subspace of the true embedding space is given
by the slope of the graph. The slope does not increase from T = 7 to T = 10,
indicating that 7 consecutive days of data suffice to define the state of
the system (a point in phase space).

\noindent Figure 6  The scaling graph for T = 29 consecutive days of data shows
very clearly an increase in the slope at low values of r, up to r E 0.065
where there is no question that the effect observed is statistically
significant, since the total number of pairs observed is then equal to
401560.05 E 200.  The higher slope Deff E 36 shows that the correlation
dimension of the coupled system is at least this large.

\noindent Figure 7  The distance between two initially nearby points as they
evolve in the pattern set is given as a function of the evolution time.
After a rapid rise which reflects the short term chaotic dynamics, this
distance follows a slowly rising slope towards the upper limit which
corresponds to the distance between two randomly chosen patterns.

\noindent Figure 8  An estimate of the short -term (chaotic) predictability is
given by the graph of the number of pairs found in a ball of fixed
radius  as a function of the projected embedding dimension, d=19T.
This number decreases exponentially with the dimension d, and the rate
of exponenial decline is a measure of predictability.

\noindent Figure 9  Another estimate of the short-term predictability is given
by
the graph of the projected attractor dimension versus the projected
embedding space dimension, D(T), in the linear region.  This measure
gives the result Tpre E 2.1, comparable to that of Figure 8.

\noindent Figure 10  The actual precipitation curve for 1992, smoothed by
10-day
averaging (solid line) is compared to the normal precipitation curve,
defined as an average over 10 years of data for dates within 10 days
of the current date (thick dotted line) and the predictions of the model
for 10-day cumulative precipitation (fine dotted line).  The shift in
the y-axis reflects the annual average which was removed. Note that 1992
was an unusual year in that the rain season begun almost two months ahead
of schedule.  This sort of phenomenon is precisely what a model should be
able to predict if it is to be useful in practical applications, such as
agriculture.  The predictions slightly exagerate the out-of-season winter
rains (first bump at the left), but predict accurately the beginning of
the rain season at day N! 60.  The other unusual phenomenon, a dry spell
followed by a recovery of rains late in the rain season, is predicted,
but with a delay of about 10 days.

\bye